\documentclass[emulateapj]{aastex61}

\usepackage{amssymb}
\usepackage{tabularx}
\usepackage{makecell}

\usepackage{lineno}
\usepackage[colorinlistoftodos]{todonotes}
\usepackage{graphicx}
\usepackage{lineno}
\usepackage{mathrsfs}
\usepackage{float}
\usepackage{array}
\usepackage{gensymb}
\newcommand{\msun}{\ensuremath{{M}_\odot}}

\begin{document}

\title{No Evidence for Chemical Abundance Variations in the
  Intermediate-age Cluster NGC 1783}

\author{Hao Zhang} 
\affiliation{Kavli Institute for Astronomy \& Astrophysics and
  Department of Astronomy, School of Physics, Peking University, Yi He
  Yuan Lu 5, Hai Dian District, Beijing 100871, China}

\author{Richard de Grijs}
\affiliation{Kavli Institute for Astronomy \& Astrophysics and
  Department of Astronomy, School of Physics, Peking University, Yi He
  Yuan Lu 5, Hai Dian District, Beijing 100871, China}
\affiliation{International Space Science Institute--Beijing, 1
  Nanertiao, Zhongguancun, Hai Dian District, Beijing 100190, China}

\author{Chengyuan Li}
\affiliation{Department of Physics and Astronomy, Macquarie
  University, Sydney, NSW 2109, Australia}

\author{Xiaohan Wu}
\affiliation{Kavli Institute for Astronomy \& Astrophysics and
  Department of Astronomy, School of Physics, Peking University, Yi He
  Yuan Lu 5, Hai Dian District, Beijing 100871, China}
\affiliation{Harvard-Smithsonian Center for Astrophysics, Harvard
  University, Cambridge, MA 02138, USA}

\begin{abstract}
We have analyzed multi-passband photometric observations, obtained
with the {\sl Hubble Space Telescope}, of the massive ($1.8 \times
10^5\ \msun$), intermediate-age (1.8 Gyr-old) Large Magellanic Cloud
star cluster NGC 1783. The morphology of the cluster's red giant
branch does not exhibit a clear broadening beyond its intrinsic width;
the observed width is consistent with that owing to photometric
uncertainties alone and independent of our photometric selection
boundaries applied to obtain our sample of red-giant stars. The color
dispersion of the cluster's red-giant stars around the best-fitting
ridgeline is $0.062 \pm 0.009$ mag, which is equivalent to the width
of $0.080 \pm 0.001$ mag derived from artificial simple stellar
population tests, that is, tests based on single-age,
single-metallicity stellar populations. NGC 1783 is comparably massive
as other star clusters that show clear evidence of multiple stellar
populations. After incorporating mass-loss recipes from its current
age of 1.8 Gyr to an age of 6 Gyr, NGC 1783 is expected to remain as
massive as some other clusters that host clear multiple populations at
these intermediate ages. If we were to assume that mass is an
important driver of multiple population formation, then NGC 1783
should have exhibited clear evidence of chemical abundance
variations. However, our results support the absence of any chemical
abundance variations in NGC 1783.
\end{abstract}


\section{Introduction}\label{intro}

Star clusters were long believed to be composed of simple stellar
population (SSPs). That is, they were thought to consist of stars that
were all formed at approximately the same time in a single starburst
event from a common progenitor giant molecular cloud. This would also
imply that all of their stars should have similar metallicities
\citep{2013MNRAS.429.1913V, 2014MNRAS.443.3594B, 2014MNRAS.441.2754C,
  2016RAA....16..179L, 2017MNRAS.472...67H}. To avoid stochastic
sampling effects, statistical studies of star clusters and their
stellar populations are often based on clusters with masses exceeding
a few $\times 10^4$--$10^5\ \msun$, thus precluding the use of open
clusters in the Milky Way or the Magellanic Clouds for such purposes.

Today, we know that the SSP approximation holds true for at least some
star clusters; after all, no evidence of chemical abundance variations
has been found among the stars populating most young massive clusters
\citep[e.g.,][]{2008AJ....136..375M, 2016MNRAS.460.1869C}. In
addition, it has been suggested that rapid star formation will eject
member stars and residual gas from young clusters
\citep{2003MNRAS.339..577B, 2007MNRAS.380.1589B, 2015MNRAS.450.2451F}.
This is supported by observations of extremely young embedded clusters
\citep{2014A&A...568A..16L} and young massive clusters
\citep{2006MNRAS.369L...9B, 2012ApJ...761..155D, 2013MNRAS.436.2852B,
  2014MNRAS.443.3594B, 2015MNRAS.448.2224C}.

However, an ever larger body of observations of old globular clusters
(GCs; with ages $> 10$ Gyr) increasingly challenges this model. For
example, studies based on small samples of bright stars in individual
clusters have revealed significant star-to-star chemical dispersions
that violate the SSP scenario \citep{1996AJ....112.1517S,
  2004ApJ...610L..25C, 2004ARA&A..42..385G,
  2009A&A...505..117C}. Photometric and spectroscopic analyses of
massive star clusters have provided additional support for the
presence of intra-cluster chemical abundance dispersions
\citep{1994ApJS...95..107W, 1996AJ....112.1517S, 2000A&A...363..159B,
  2007ApJ...661L..53P, 2008A&A...490..625M, 2009ApJ...697L..58A,
  2013ApJ...775...15P}.

Given that most young massive clusters appear to be clean SSPs, while
many old GCs show evidence of containing multiple stellar populations
(MSPs), the key questions of interest become at which age the MSP
phenomenon first makes an appearance, and whether age is indeed the
crucial parameter driving the emergence of MSPs. That is, is there a
critical age at which the transition from SSPs to MSPs occurs? Could
mass be an alternative parameter? The latter scenario may not be as
far-fetched as it might appear at first sight: indeed, if MSPs
originate primarily from colliding asymptotic giant-branch (AGB)
stellar winds, one would expect MSPs to only form in clusters that are
sufficiently massive to retain the AGB ejecta in their gravitational
potential wells.

In the context of cluster formation through an initial starburst
\citep[e.g.,][]{2013MNRAS.429.1913V, 2014MNRAS.443.3594B,
  2014MNRAS.441.2754C, 2017MNRAS.472...67H}, the presence of MSPs
would be primarily associated with the formation mechanism of
second-generation stars. Most related scenarios suggest that these
second-generation stars would be characterized by broad distributions
of chemical abundances, because they formed from gas that was polluted
by the ejecta from first-generation stars. Possible polluters include
rapidly rotating massive stars \citep{2013MmSAI..84..158C},
supermassive stars \citep{2014MNRAS.437L..21D}, massive interacting
binary systems \citep{2015MNRAS.449.3333B}, young AGB stars
\citep{2008MNRAS.391..825D}, and evolving low-mass binary systems
\citep{2014ApJ...789...88J}.

Alternative scenarios have been proposed which suggest external
accretion, including accretion from the interstellar medium
\citep{2009MNRAS.394..124B, 2009MNRAS.397..488P, 2016Natur.529..502L},
accretion of AGB ejecta \citep{2017MNRAS.467.1857B}, and cluster
mergers \citep{2017MNRAS.472...67H}. Unfortunately, the star cluster
population in the Milky Way has an extreme age distribution
\citep{1996AJ....112.1487H, 2005ApJS..161..304M}, given that most GCs
are older than 10 Gyr, while only a handful of similarly massive
clusters are younger than a few $\times 10^8$ yr. Therefore, the
massive cluster population in the Milky Way cannot be used to test
whether MSPs emerge with increasing cluster age. Similarly, in the
Large Magellanic Cloud (LMC), a well-known `age gap' exists in the
cluster age distribution, with a dearth of clusters in the 2--6 Gyr
age range \citep{1997AJ....114.1920G, 2010MNRAS.404.1625B}.

Fortunately, the combined sample of massive clusters in the LMC and
the Small Magellanic Cloud (SMC) includes a number of objects in the
2--6 Gyr range \citep{2009A&A...497..755M}. Specifically, six star
clusters from the \citealt{2005ApJS..161..304M} catalog fall in this
age range, including NGC 2121, Hodge 4, NGC 2155, NGC 2193, SL 663,
and Kron 3. The youngest star cluster that has been found to show
evidence of MSPs is NGC 1978 \citep{2018MNRAS.473.2688M}. It has an
age close to 2 Gyr. Therefore, assuming that age is a key driver of
the MSP phenomenon, its onset appears to coincide with an age of
around 2 Gyr.

Chemical abundance variations provide important evidence for
(primarily old) stellar population studies
\citep{2004ARA&A..42..385G}. Spectroscopic observations of the
presence of chemical abundance variations in bright giant stars have
provided evidence of Na--O \citep{2004ApJ...610L..25C,
  2004ARA&A..42..385G, 2009A&A...505..117C} and Mg--Al
anticorrelations \citep{1996AJ....112.1517S}. Multimodal stellar
distributions in color--magnitude diagrams (CMDs) are observed in some
star clusters. Indeed, many star clusters exhibit two or more parallel
sequences in their CMDs \citep{2008A&A...490..625M,
  2015AJ....149...91P}, which---for old GCs---have been linked to
variations in chemical abundances. Significantly broadened sequences
in cluster CMDs could also be indicative of chemical variations,
although at the present time conclusive studies are lacking at
intermediate ages.

The most common elements that may give rise to multiple parallel
sequences in a cluster's CMD and which can be easily traced
spectroscopically include He and N
\citep{2004ARA&A..42..385G}. Variations in He abundances affect
stellar optical luminosities \citep{2004ARA&A..42..385G}, while N
abundances determine the ultraviolet (UV) luminosities
\citep{2012ApJ...744...58M, 2015MNRAS.447..927M}. Thus, by combining
photometry obtained in optical and UV passbands, one could potentially
constrain chemical abundance variations in star cluster CMDs.

In this paper, we aim at constraining the maximum chemical abundance
spread in an intermediate-age cluster with an age close to the lower
boundary of the 2--6 Gyr age range of interest. We considered the
suitability of a number of potential target clusters and eventually
settled on a careful analysis of NGC 1783. This LMC cluster has an age
of approximately 1.8 Gyr (see Section \ref{isochrone}) and a mass of
$1.8 \times 10^5\ \msun$ \citep{2013MNRAS.430..676B}. Here, we
investigate NGC 1783 using the same methodological approach as
\cite{2017MNRAS.468.3150M, 2018MNRAS.473.2688M}. We compare the
morphology of the cluster's red-giant branch (RGB) with that of an
artificial SSP. As we will see below, the width of the RGB in NGC 1783
is consistent with that expected from an SSP, which implies that the
cluster's chemical abundances are essentially single-valued and
homogeneous.

We will particularly compare our results with those of
\citep{2018MNRAS.473.2688M}. The latter authors have pursued similar
research questions based on a large sample of intermediate-age
clusters, including NGC 1783. Although their method has a solid
mathematical foundation, relying on Gaussian mixture modeling and
Akaike information criteria, it is unclear whether their Gaussian
peaks have widths that are consistent with the prevailing photometric
errors. Our research goes one step beyond their approach by
implementing a careful comparison using artificial stars. We point out
that our approach is more realistic and based on physical
considerations, while fully taking into account the relevant
photometric uncertainties. This will allow us to constrain the
intrinsic width of the RGB as well as quantify any evidence of
broadening, if present.

This paper is organized as follows. In Section 2, we discuss the
observations, as well as our data reduction and analysis
procedures. In Sections 3 and 4, we present our results and discuss
their astrophysical implications. Section 5 summarizes and concludes
the paper.

\section{Data Reduction}

\subsection{Data and Photometry}\label{reduction}

We used {\sl Hubble Space Telescope} ({\sl HST}) Advanced Camera for
Surveys (ACS)/Wide Field Camera (WFC) and {\sl HST} Ultraviolet
Visible channel (UVIS)/Wide Field Camera 3 (WFC3) archival data
obtained from the {\sl HST} Legacy Archive. Specifically, we
downloaded images obtained through the F336W, F435W, F555W, and F814W
filters. For each photometric band, except for the F336W filter, we
obtained long- and short-exposure images in order to ameliorate
saturation effects. For F336W, we only obtained a long-exposure image;
however, this latter image does not suffer from saturation. Images
taken with the {\sl HST} ACS/WFC (Proposal ID: GO-10595; Principal
Investigator, PI: Goudfrooij) include long- and short-exposure images
in the F435W, F555W, and F814W bands. The long exposure times were,
respectively, 770 s, 720 s, and 688 s, while the corresponding short
exposure times were 90 s, 40 s, and 8 s. Our long-exposure F336W image
was taken with the {\sl HST} UVIS/WFC3 (Proposal ID: GO-12257; PI:
Girardi), with an exposure time of 3580 s.

We used two independent photometric software packages, i.e., Iraf/{\sc
  daophot} \citep{1987PASP...99..191S} and {\sc Dolphot}
\citep{2000PASP..112.1383D}, to perform point-spread-function (PSF)
photometry on the NGC 1783 images. We compared the outputs of the two
packages, concluding that both yield internally consistent photometric
results \citep[see also][their Fig. 5]{2016ApJ...826L..14W}. Our
subsequent analysis will be based on the {\sc Dolphot} output in the
Vegamag photometric system.

We proceeded by following the recommendations in the {\sc Dolphot}
manual:\footnote{http://purcell.as.arizona.edu/dolphot/}

\begin{itemize}

\item Mask bad pixels using the {\tt ACSMASK/WFC3MASK} command;

\item Split the images according to the chips used for the
  observations by application of the {\tt SPLITGROUPS} command;

\item Calculate the sky brightness level using the {\tt CALCSKY}
  command;

\item Proceed to perform PSF photometry using the {\tt DOLPHOT}
  command. For the ACS/WFC bands (F435W, F555W, and F814W), we
  performed our photometric approach on all three long-exposure {\tt
    {\_}flt} images simultaneously, using the drizzled ({\tt {\_}drz})
  F435W image as our reference image. We adopted this latter image as
  our reference since it has the longest exposure time, although we
  could have taken any of the long-exposure images without suffering
  from selection biases. We performed PSF photometry separately on the
  long-exposure F336W {\tt {\_}flt} image.

\item Repeat the PSF photometry for the short exposure {\tt {\_}flt}
  images of the ACS/WFC bands (F435W, F555W, and F814W) with exactly
  the same approach and the same reference image.

\item Combine the catalogs obtained from the different chips.

\end{itemize}

We only selected objects characterized by sharpness parameters in the
output catalog within the range $[-0.3, 0.3]$. The combined F435W,
F555W, and F814W long-exposure catalog contained 36,091 stars; the
short-exposure catalog included 12,782 stars. From the long-exposure
F336W image, we retrieved 32,792 objects.

\begin{figure}[h]
\centering
\includegraphics[width=0.4\textwidth]{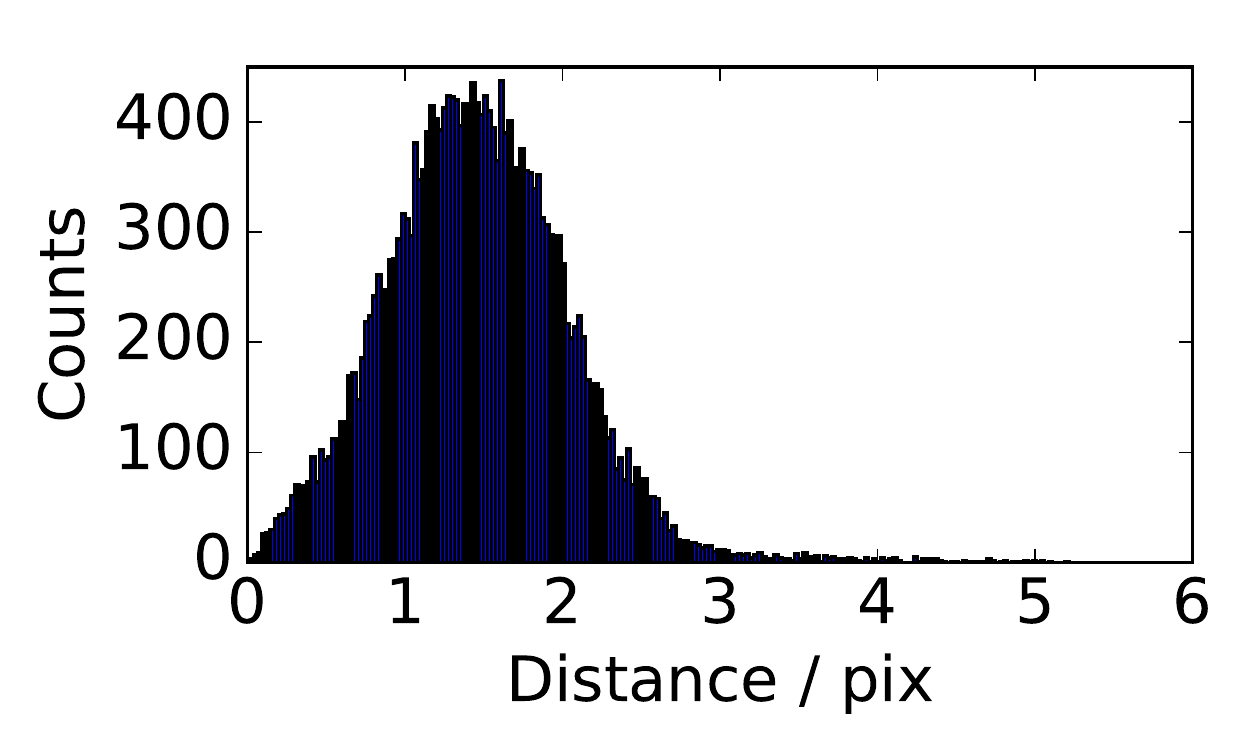}
\caption{Histogram showing the positional mismatch distribution
  between the F336W catalog and the combined F435W, F555W, and F814W
  photometric catalog.}
\label{pos_mismatch}
\end{figure}

To cross-identify sources between the catalogs and filters, we applied
a critical (maximum) distance of $0.3''$ ($\sim$6 pixels). The cross
matching was done in R.A.--Dec. space, so that any effects associated
with image translation and rotation were taken into account
properly. Inspection of the histogram of position mismatches between
the F336W catalog on the one hand and the combined catalog resulting
from the F435W, F555W, and F814W images (which had already been
internally aligned to subpixel accuracy) on the other shows that the
distribution of positional mismatches peaks around 1.5 pixels: see
Fig. \ref{pos_mismatch}.

\begin{figure}[h]
\centering
\includegraphics[width=0.4\textwidth]{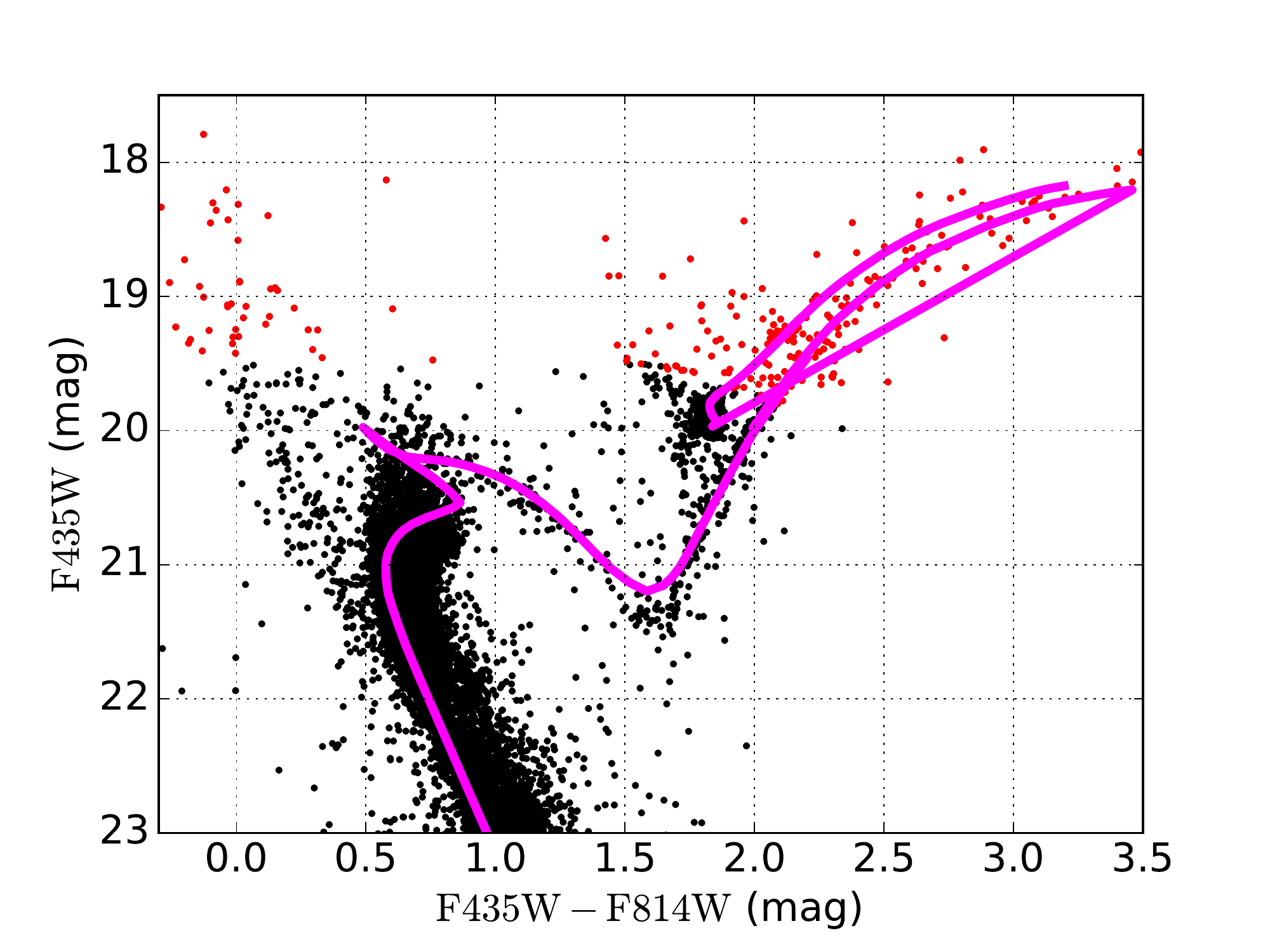}
\caption{NGC 1783 CMD and best-fitting PARSEC isochrone
  \citep{2012MNRAS.427..127B}. Black (red) points: stars fainter
  (brighter) than the saturation level. The physical parameters
  associated with the best-fitting isochrone are: Age $t = 1.8$ Gyr,
  distance modulus $(m - M)_0 = 18.46$ mag, and metallicity $Z =
  0.007$.}
\label{best1.8}
\end{figure}

We started by matching the catalogs resulting from the different
exposure times. For our short-exposure images, we only adopted those
stars characterized by $m_{\rm F435W} < 19.5$ mag or $m_{\rm F555W} <
18.8$ mag (a total of 454 objects), since fainter objects are not
saturated and, hence, the longer exposure time images are expected to
yield more reliable photometry: see Fig. \ref{best1.8} for an
example. For unsaturated objects matched in both catalogs, we adopted
the long-exposure photometry. Unmatched stars from both the short- and
long-exposure images were combined to construct our final catalog,
containing 36,206 stars.

We finally cross-identified objects found in different filters using
the same procedure. We only retained objects matched in all
filters. The matched star catalog with magnitudes in all four
photometric bands contains 24,425 stars.

\subsection{Differential Reddening Correction}

Differential reddening corrections were applied to properly deal with
the potentially spatially variable foreground reddening. An example to
underscore this importance is provided by \citep[][their
  Fig. 8]{2012A&A...540A..16M}. Their results show clear differences
in the CMDs---particularly in the widths of the main sequences and
RGBs---of several clusters before and after having applied these
corrections.

Next, we needed to correct our photometry for the effects of
differential reddening. Again, we followed the procedure proposed by
\cite[][their Section 3.1]{2012A&A...540A..16M}:

\begin{itemize}

\item Rotate the CMD such that the abscissa axis becomes the direction
  of the differential reddening. The rotation angle can be calculated
  from the ratio of the extinction effects in the two filters used to
  construct the CMD;

\item Select all stars from a smooth region in the CMD as reference
  stars. We selected the upper main-sequence region for this purpose
  (specific magnitude and color limits depend on the passband
  combination used). We proceeded to fit a central fiducial curve to
  this smooth region. Note that we only measure relative reddening
  values, which hence might be negative with respect to the value
  pertaining to the reference curve. We assume the deviation of each
  point from the central curve to be fully determined by differential
  reddening, although---particularly along the main sequence---the
  presence of binary systems may also contribute to the
  deviations. However, since we are predominantly interested in the
  cluster's RGB morphology, this does not affect our results;

\item For any star in the catalog, we determine the associated
  reddening value by averaging the reddening values of its spatially
  $N$-nearest reference stars. We adopted $N = 40$
  \citep[cf.][]{2012A&A...540A..16M}.
  
\end{itemize}

\begin{figure}[h]
\centering
\includegraphics[width=1.0\textwidth]{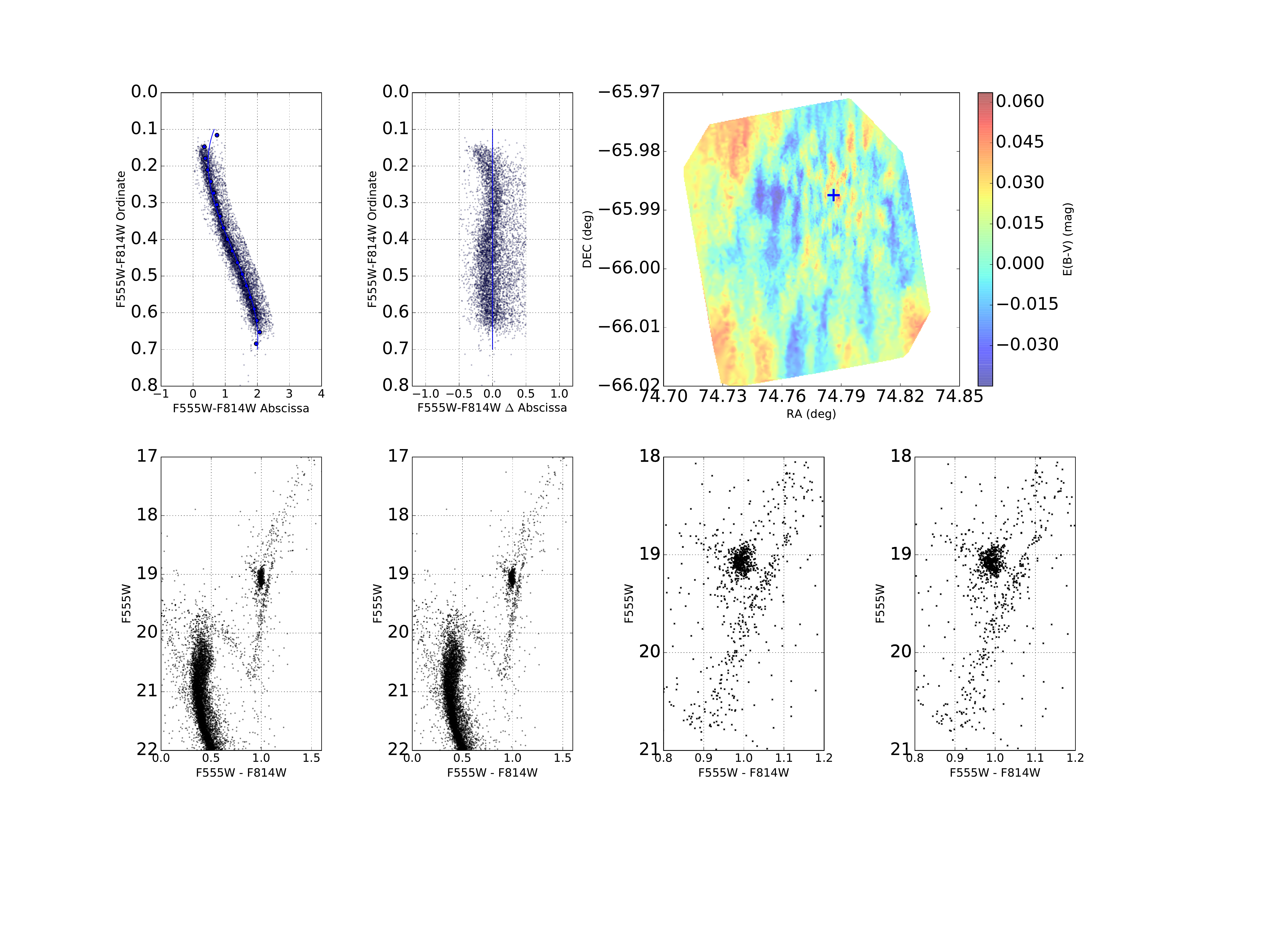}
\caption{(Top row, left and middle) Fiducial curve fitting and
  determination of the reddening values for reference stars in
  abscissa--ordinate space. (Top row, right) Spatial differential
  reddening map, where the cluster center is marked by a plus
  sign. The color coding represents the $E(B-V)$ values. (Bottom row,
  left) CMDs before and after correction. (Bottom row, right) As the
  two left-hand panels, but focusing on the lower-RGB region.}
\label{DRC}
\end{figure}

Since we have access to observations in four filters, thus allowing
six passband combinations, we repeated this analysis for each of the
six possible pairs of passbands. Hence, we obtained three corrected
catalogs for each band; for instance, the F336W catalog was corrected
using the combinations $m_{\rm F336W} - m_{\rm F435W}$, $m_{\rm F336W}
- m_{\rm F555W}$, and $m_{\rm F336W} - m_{\rm F814W}$. For each star,
the average corrected magnitude from all three corrected catalogs was
taken as the final, corrected magnitude. As an example, Fig. \ref{DRC}
shows the differential reddening map for the $m_{\rm F555W} - m_{\rm
  F814W}$ filter combination \citep[see also][their
  Fig. 5]{2012A&A...540A..16M}. This differential reddening map was
applied to the uncorrected photometry to obtain the de-reddened CMD.

\subsection{Selection of Red Giant Stars}

\begin{figure}[h]
\centering
\includegraphics[width=0.7\textwidth]{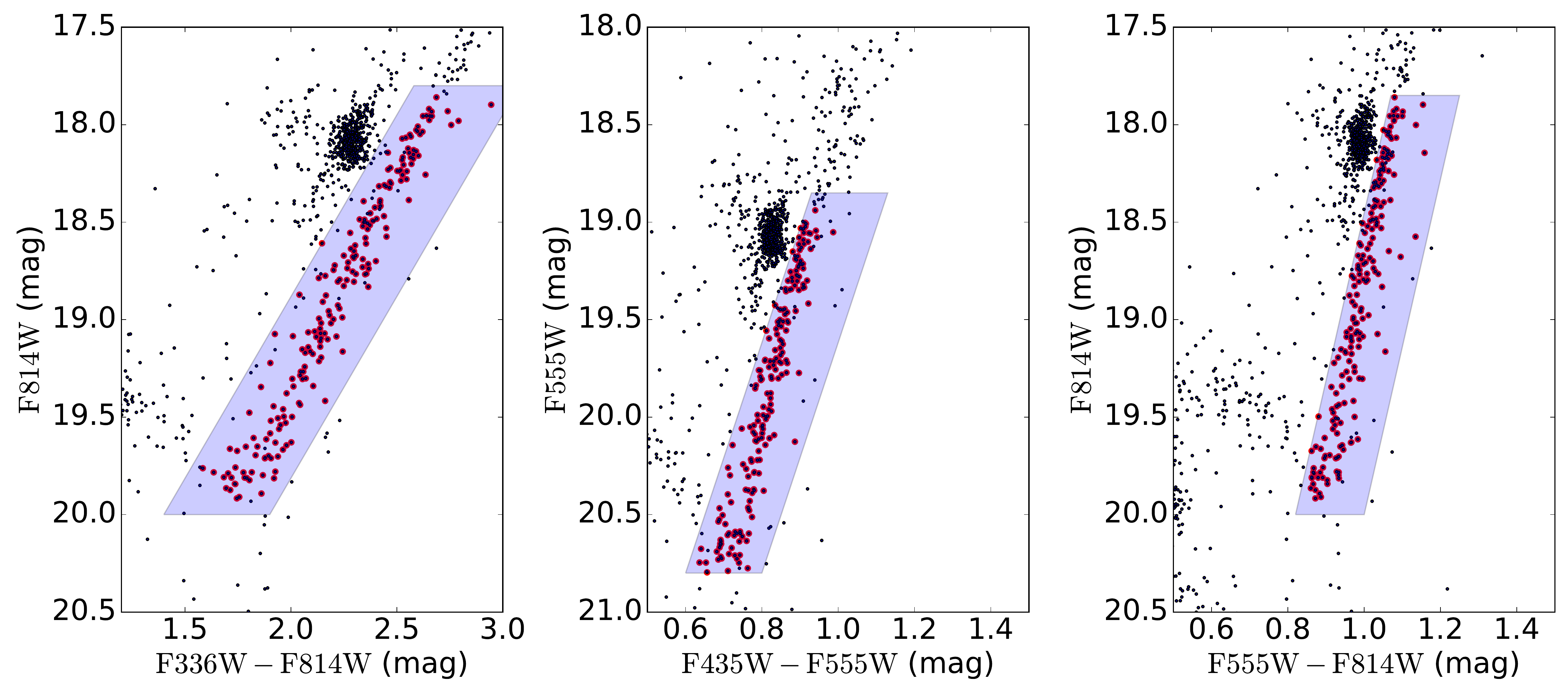}
\caption{RGB sample selection. Only stars simultaneously found in all
  three parallelograms were considered for selection.}
\label{RGB_selection}
\end{figure}

Figure \ref{RGB_selection} shows how we selected a sample containing
lower-RGB stars for subsequent analysis. We only selected stars of
similar luminosity or fainter than the cluster's red clump (RC);
selection of brighter RGB stars would open up our analysis to
contamination by AGB stars. We applied simultaneous
parallelogram-based selection in the $m_{\rm F336W}$ versus $m_{\rm
  F336W}-m_{\rm F814W}$, $m_{\rm F435W}$ versus $m_{\rm F435W}-m_{\rm
  F555W}$, and $m_{\rm F555W}$ versus $m_{\rm F555W}-m_{\rm F814W}$
CMDs, only retaining RGB stars that were found in all three
parallelograms. The relevant magnitude ranges pertaining to our
selection of RGB stars were $17.8 < m_{\rm F814W} < 20.0$ mag and
$18.85 < m_{\rm F555W} < 20.8$ mag, so that our RGB sample stars cover
the range from the bottom of the RGB up to the bright limit of the
RC. Although we did not implemented such magnitude cuts in the F336W
and F435W bands, the corresponding ranges in these passbands are $20.5
< m_{\rm F336W} < 21.75$ mag and $19.85 < m_{\rm F435W} < 21.55$
mag. These boundaries also determine magnitude ranges from the bottom
of the RGB to the RC's upper limit.

The equivalent color ranges used for our RGB selection were $\Delta
(m_{\rm F336W}-m_{\rm F814W}) = 0.50$ mag, $\Delta (m_{\rm
  F435W}-m_{\rm F555W}) = 0.20$ mag, and $\Delta (m_{\rm F555W}-m_{\rm
  F814W}) = 0.18$ mag, based on RGB color dispersions of 0.076, 0.025,
and 0.028 mag, respectively. We selected the precise color boundaries
such that the RGB was well separated from the RC; any rejection of RGB
stars owing to photometric errors was minimized by the wide color
selection ranges, corresponding to more than five times the observed
RGB sample dispersions.

Our lower RGB sample thus selected includes 196 stars; 64 stars were
rejected because they were only included in one or two of our
selection parallelograms.

\subsection{Field Star Decontamination}

Before quantifying the width of the RGB, we must ensure that our RGB
sample is not contaminated by field RGB stars. The cluster's center is
located at $\mathrm{R.A.}_{\rm J2000}\ =\ 4^{\rm h}59^{\rm m}8.68^{\rm
  s},\ \mathrm{Dec.}_{\rm J2000} = -65^{\circ}59'14.82''$
\citep{2016RAA....16..179L}. We adopted a cluster radius of $75''$,
i.e., three times the cluster's core radius for stars with $m_{\rm
  F435W} <23$ mag \citep{2016Natur.529..502L}.

\begin{figure}[h]
\centering
\includegraphics[width=0.4\textwidth]{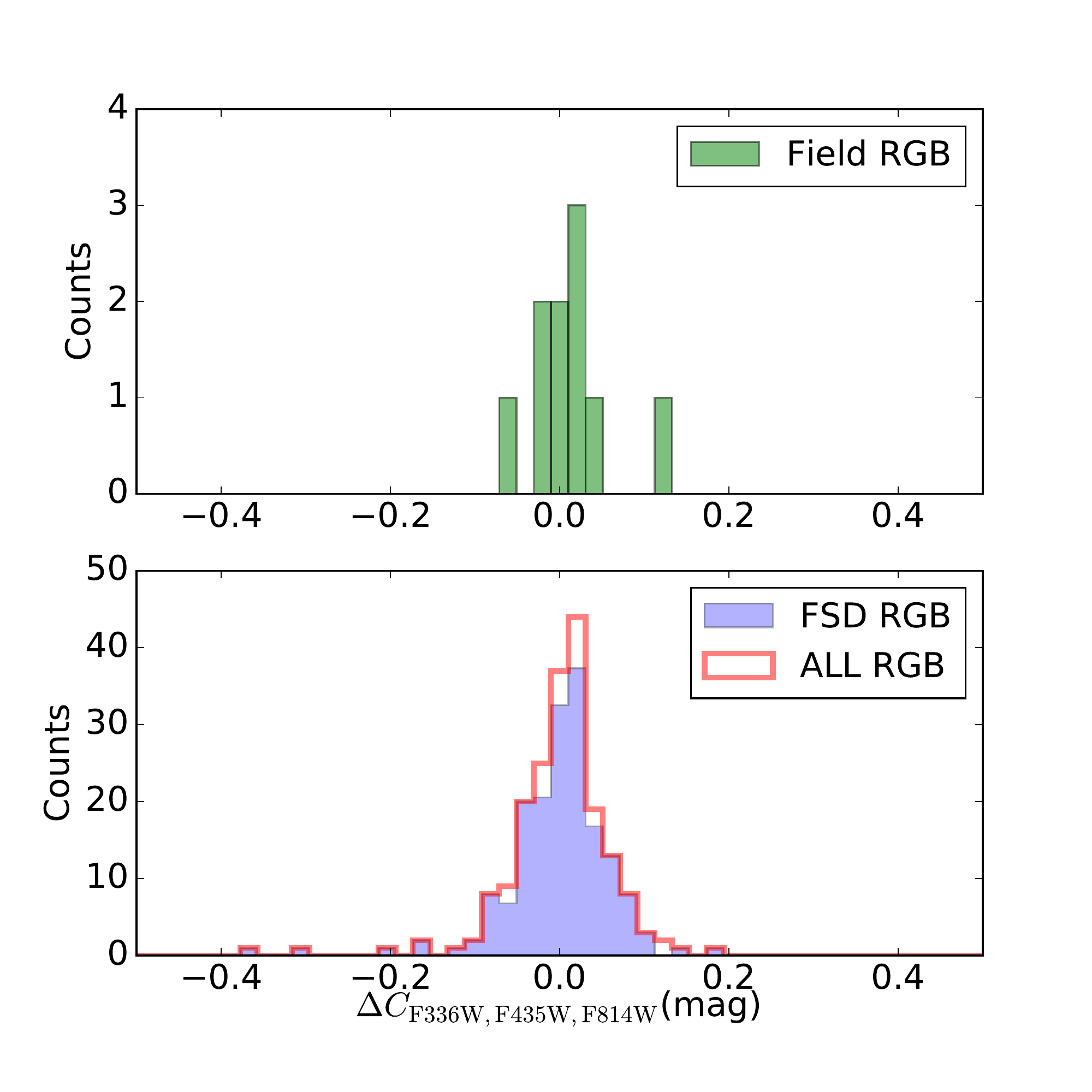}
\caption{Field-star decontamination and its effect on the
  $C_{\mathrm{index}}$ histogram. (Top) $C_{\mathrm{index}}$ histogram
  of RGB stars in the reference field. (Bottom) Normalized
  $C_{\mathrm{index}}$ histograms of all RGB stars and of the
  field-star decontaminated (FSD) RGB stars.}
\label{RGB_FSD}
\end{figure}

Our field-star decontamination procedure proceeded on the basis of the
following steps (see Fig. \ref{RGB_FSD}):

\begin{itemize}

\item We determined a field-dominated region outside the cluster's
  boundary region;
\item We then applied the same RGB selection criteria to the cluster
  and the field CMDs to obtain the equivalent field RGB sample;
\item We finally statistically subtracted a field component from the
  distribution of the cluster region RGB sample, taking into account
  the different areas used for the cluster and field regions. The
  ratio of the cluster and field regions used was 2.223.

\end{itemize}

We emphasize that, rather than directly assigning `field star’ status
to some of our lower-RGB sample and removing them, instead we subtract
the $C_{\mathrm{index}}$ distribution (which is described in detail in
Section \ref{AST} below) of our field region from the same
distribution retrieved from the cluster region.

\section{Results}

\subsection{Isochrone Fitting}\label{isochrone}

We adopted the average extinction, $A_V = 0.06$ mag, and $(m - M)_0 =
18.46$ mag from \citealt{2016Natur.529..502L}. We used these values to
explore the performance of a set of PARSEC stellar isochrones
\citep{2012MNRAS.427..127B}, adjusting their metallicity $Z$ and age
$t$ to visually search for the best-matching isochrone. We obtained as
best-fitting parameters $t = 1.8 \pm 0.1$ Gyr and $Z = 0.007 \pm
0.001$. The resulting isochrone is overplotted on the CMD in
Fig. \ref{best1.8}.

\cite{2011ApJ...737....3G} obtained $t = 1.7 \pm 0.1$ Gyr, while
\cite{2007MNRAS.379..151M} determined $t = 1.4 \pm 0.2$ Gyr, $A_V =
0.13$ mag, and $(m - M)_0 = 18.45$ mag. \cite{2013MNRAS.430.2774R}
obtained $A_V = 0.22$ mag, $(m - M)_0 = 18.57$ mag, and $Z =
0.0065$. Finally, \citealt{2016Natur.529..502L} recommended $t = 1.4$
Gyr, $Z = 0.008$, and $A_V = 0.06$ mag. With respect to the latter
determination, we believe that our best-fitting metallicity of
$Z=0.007$ is more accurate; we carefully examined the match of our
best-fitting isochrone to the cluster's RC. For the best-fitting age
of $t = 1.8$ Gyr, the resulting metallicity of $Z = 0.007$ is clearly
the best fit.

Second, the age derived by \cite{2013MNRAS.430.2774R} was relatively
poorly constrained, since it was based on obtaining the optimal fit to
the morphology of the extended main-sequence turn-off
\citep{2009A&A...497..755M}. The age difference of 400 Myr implied by
our best fit here likely reflects a more accurate age determination,
which is constrained by the overall morphology of the cluster CMD,
including that of the post-main-sequence evolutionary phases.

\subsection{Artificial Star Tests}\label{AST}

We used artificial star tests to examine whether the full width of the
RGB can be explained solely by the photometric uncertainties. Using
the best-fitting isochrone and the spatial distribution of the RGB
stars, we created an artificial input stellar catalog containing
78,400 stars. We subsequently reran the {\sc Dolphot} photometry using
the {\tt -fakestar} option. Of the 78,400 input stars, we recovered
68,800 objects. The 78,400 artificial stars were inserted in 400 sets
of 196 objects at a time so as to avoid adding too many artificial
stars to the cluster image simultaneously. If we were to overload the
science image with artificial stars, unrealistically high blending
ratios and background levels would result.

We examined the observed and artificial RGB sequences by plotting
their magnitudes versus their color index; the latter is the color
defined by the objects' magnitudes in three bands, i.e.,
$C_{\mathrm{F336W, F435W, F814W}} = (m_{\rm F336W} - m_{\rm F435W}) -
(m_{\rm F435W} - m_{\rm F814W})$. Second-generation stars are expected
to have higher N and He abundances, and lower C abundances, than
first-generation stars. Their $(m_{\rm F435W} - m_{\rm F814W})$ colors
will hence become bluer because of the increased temperatures caused
by more abundant He, while the $(m_{\rm F336W} - m_{\rm F435W})$
colors will become redder owing to a nitrogen absorption band found
within the F336W filter bandpass, which is enhanced by the higher N
abundance. Therefore, these differences between both generations of
stars add up in the $C_{\mathrm{F336W, F435W, F814W}}$ combination,
which is thus a good choice to probe chemical abundance variations
associated with MSPs. Such double-peak features in the
$C_{\mathrm{index}}$ distributions of the RGB have been discovered in,
e.g., NGC 121 \citep{2017MNRAS.465.4159N} and NGC 1978
\citep{2018MNRAS.473.2688M}---based on the use of similar instruments
and exposure times as done in this paper---which has been interpreted
as clear evidence of dual chemical abundance components. This
indicates that application of the $C_{\mathrm{index}}$ is indeed a
suitable approach to detect chemical abundance variations and that the
absence of a $C_{\mathrm{index}}$ dispersion implies a lack of
evidence of any abundance variations in one's cluster of interest.

To generate the input artificial stellar catalog, we proceeded as follows:

\begin{itemize}
\item We assigned random coordinates to the artificial stars based on
  a random selection of 196 lower-RGB stars in our observed sample;
\item To avoid spatial overlaps with any of the observed lower-RGB
  stars, we added random perturbations to these coordinates. We
  adopted perturbations equivalent to half the distance of any of the
  selected lower-RGB stars to their (spatially) closest lower-RGB star
  in our sample.
\item We assigned magnitudes to the artificial stars based on the
  closest-matching segment of the best-fitting PARSEC isochrone. We
  interpolated this segment uniformly (i.e., we adopted a curve in the
  four-dimensional space defined by the F336W, F435W, F555W, and F814W
  magnitudes), for the entire artificial-star catalog.
\item Finally, we matched the coordinates randomly with the resulting
  magnitudes to complete our input artificial RGB star catalog.
\end{itemize}

We next used {\sc Dolphot} to obtain PSF photometry for the artificial
stars we just inserted into the science images and generated the
corresponding output photometric catalog. This procedure is fully
equivalent to that adopted for the real stars in Section
\ref{reduction}.

The results of our artificial-star analysis can also be used to
estimate the completeness levels of our observations. We recovered
68,880 of the 78,400 input stars, so the completeness level of our RGB
sample is approximately 88\%. Figure \ref{best1.8} shows that the
photometric detection limit is significantly fainter than $m_{\rm
  F435W} = 23$ mag. Our RGB sample is significantly too bright to be
affected by the detection limit, but at the RGB level incompleteness
may be caused by crowding effects (blending) instead.

\subsection{Broadening of the Red Giant Branch} \label{RGB}

\begin{figure}[h]
\centering
\includegraphics[width=0.4\textwidth]{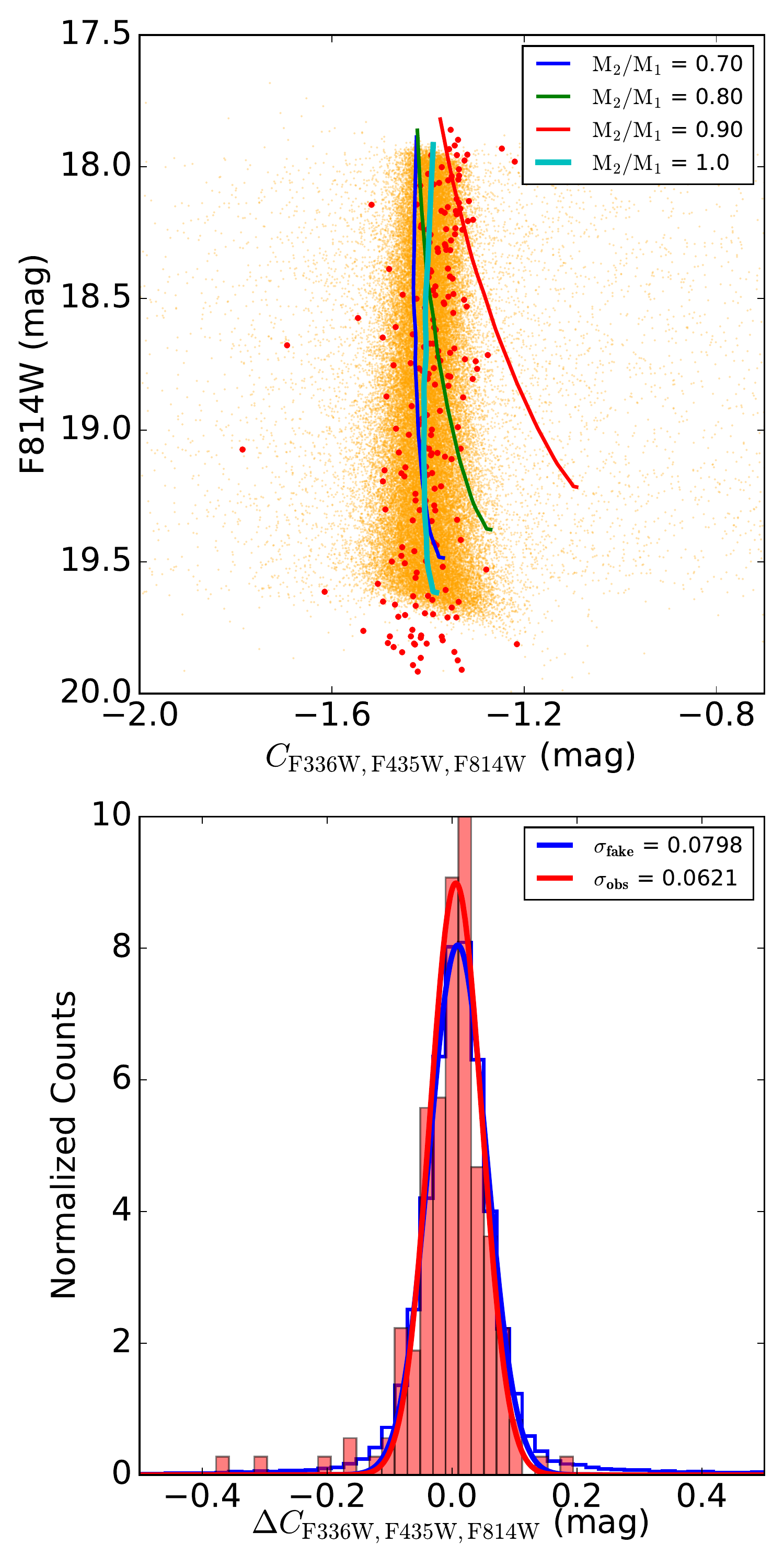}
\caption{(Top) Comparison of the artificial stellar sample (orange)
  with the observed RGB sample in the $m_{\rm F814W}$ versus
  $C_{\mathrm{F336W, F435W, F814W}}$ diagram. A set of binary
  sequences, adopting different secondary-to-primary mass ratios (see
  the legend) are also shown. The two stars at the far left cannot be
  binary systems, but they may be field stars. (Bottom) Normalized
  histograms showing the deviations from the ridgelines of both the
  observed RGBs (196 stars) and the artificial stars (68,880 stars
  representing an SSP).}
\label{ASTseq_dUUIB}
\end{figure}

We will now examine if the RGB of NGC 1783 is consistent with a single
metallicity using a similar approach as \cite{2017MNRAS.464...94N}. We
start by comparing the corresponding artificial stellar population and
the observed RGB population in the $m_{\rm F814W}$ versus
$C_{\mathrm{F336W, F435W, F814W}}$ diagram: see
Fig. \ref{ASTseq_dUUIB}, where we have also added a small translation
in color to the artificial stars' RGB ridgeline to match it to the
ridgeline of the observed RGB stars. The majority of the artificial
stars is concentrated well within the observed RGB width, while only a
small fraction is characterized by a wide color spread caused by
blending. The tight, observed RGB sequence is visually consistent with
the single-component distribution of the artificial stars.

\begin{figure}[h]
\centering
\includegraphics[width=0.4\textwidth]{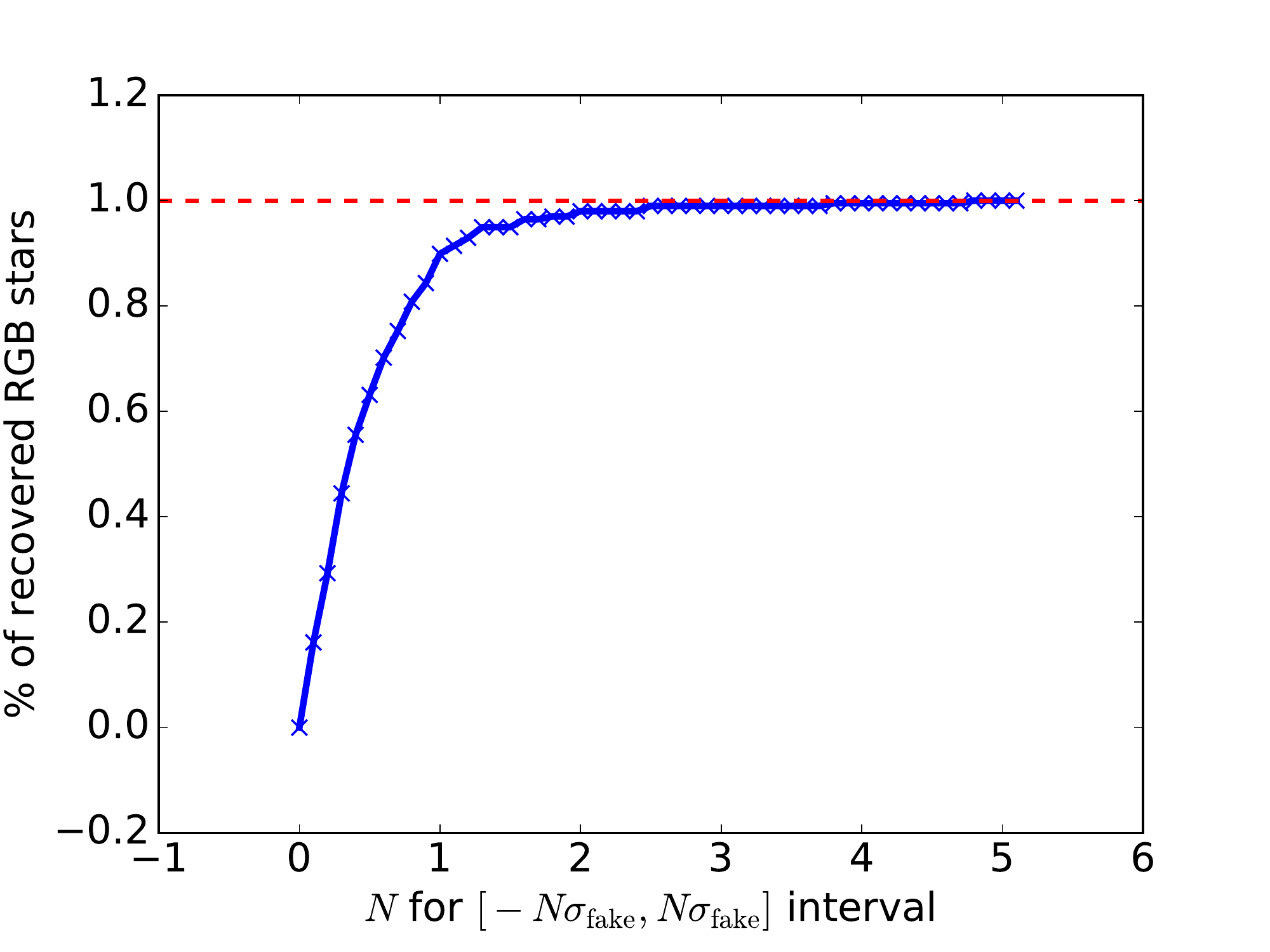}
\caption{Fraction of the observed RGB stars recovered as a function of
  the interval covered with respect to the $\Delta C_{\mathrm{F336W,
      F435W, F814W}}$ distribution; $\sigma_{\rm fake}$ is the
  dispersion of the $C_{\mathrm{F336W, F435W, F814W}}$ Gaussian
  distribution of the artificial stars.}
\label{percent_recover}
\end{figure}

We next overplotted the observed RGB broadening histogram (i.e., the
deviation from the ridgeline, $\Delta C_{\mathrm{F336W, F435W,
    F814W}}$) and that defined by the artificial stars, both
normalized. We fitted Gaussian curves to both distributions. We used a
bootstrapping method, repeating the procedure 1000 times, to measure
the Gaussian widths and their associated uncertainties. The
best-fitting Gaussian standard deviations corresponding the observed
and artificial RGB $C_{\mathrm{index}}$ distributions are $\sigma_{\rm
  obs} = 0.062 \pm 0.009$ mag and $\sigma_{\rm fake} = 0.080 \pm
0.001$ mag, respectively. Both distributions have similar widths (see
the bottom panel of Fig. \ref{ASTseq_dUUIB}). Next, we investigated
the fraction of observed RGB stars that could be reproduced by the
artificial stellar distribution in Fig. \ref{percent_recover}. We
found that the $[-1\sigma_{\rm fake},\ 1\sigma_{\rm fake}]$ interval
contains 84\% of the observed stars, while the $[-3\sigma_{\rm
    fake},\ 3\sigma_{\rm fake}]$ interval includes 99\% of the
observed stars.

Next, we adjusted our RGB selection criteria to test whether our
initial selection had excluded real RGB stars and, if so, how that
might affect our results, particularly as regards the width of the
RGB. We shifted the upper edges of our RGB selection box slightly to
bluer colors in $(m_{\rm F336W}-m_{\rm F814W})$ and $(m_{\rm
  F555W}-m_{\rm F814W})$ but not in $(m_{\rm F435W}-m_{\rm
  F555W})$. This adjusted choice leads to the inclusion of a number of
RGB stars that are located close to the selection boundary, while
retaining the color selection in $(m_{\rm F435W}-m_{\rm F555W})$ will
still lead to rejection of RC stars. The RGB sample resulting from
this updated selection run includes 207 stars. We repeated all of our
subsequent analysis and found a new $C_{\mathrm{index}}$ value of
0.063 mag, which is almost the same as our initial result,
$C_{\mathrm{index}} = 0.062$ mag. We therefore conclude that the
potential loss of a small number of genuine RGB stars by our initial
selection method does not significantly influence our measurement of
the cluster's RGB broadening, nor our final results.

\section{Discussion}

Our results show that the artificial stellar population is
characterized by a similar dispersion in $C_{\mathrm{F336W, F435W,
    F814W}}$ as the observed RGB sequence of NGC 1783. This
demonstrates that the photometric uncertainties fully account for the
RGB's broadening in NGC 1783.

The color index used, $C_{\mathrm{F336W, F435W, F814W}}$, is sensitive
to the stellar N and He abundances, which are two of the most
representative elements that exhibit large abundance variations in
star clusters \citep{2004ARA&A..42..385G}. The consistency of the
observed RGB's width with that defined by our artificial SSP therefore
indicates that our RGB sample is best represented by single-valued N
and He abundances.

There are a small number of outliers in the observed RGB distribution,
in particular two stars at the far left (see the top panel of
Fig.~\ref{ASTseq_dUUIB}). These objects are unlikely statistical
fluctuations, given that they deviate by $>5\sigma$ from the Gaussian
peak. Alternatively, they might be field stars that have not been
properly removed by our field-star decontamination algorithm. If they
were cluster members, they would need to have unrealistically extreme
metallicities. Note that these two data points appear both in the
scatter diagram in the top panel of Fig. \ref{ASTseq_dUUIB} and in the
observed histogram in the figure's bottom panel. We suspect that these
two stars might result from stochasticity in the field's RGB sample
(see Fig. \ref{RGB_FSD}) rather than from random effects associated
with our field-star decontamination.

We explored whether these objects might be binary systems instead. If
we assume that the primary components ($M_1$) of such binary systems
are genuine RGB stars, the secondary components ($M_2$) must have
lower masses. We tested the appearance of binary systems with
different mass ratios, $M_2/M_1$. For very low-mass companions, the
change in color is almost negligible, while for equal-mass binaries no
color difference is expected compared with the single-star case. The
top panel of Fig. \ref{ASTseq_dUUIB} includes binary sequences for
$M_2/M_1 = 0.7, 0.8$, and 0.9. It is clear that the outliers at the
far left cannot be binary systems. 

An intermediate-age cluster with an age of 1.8 Gyr and a mass of $1.8
\times 10^5\ \msun$, NGC 1783 is comparably massive as other star
clusters that show clear evidence of MSPs
\citep[e.g.,][]{2018MNRAS.473.2688M}. If we assume that mass is an
important driver of MSP formation, then NGC 1783 should also have
shown clear evidence of chemical abundance variations. However, we
have found no such evidence in NGC 1783, which challenges the idea
that mass may be a primary driver of MSP formation. The similarity of
the observed RGB width in NGC 1783 with that expected for an SSP
suggests the absence of chemical abundance variations. When combined
with similar results for NGC 419, NGC 1806, and NGC 1846
\citep{2017MNRAS.468.3150M, 2018MNRAS.473.2688M}, this is indeed
inconsistent with mass being a primary driver.

\begin{figure}[h]
\centering
\includegraphics[width=0.4\textwidth]{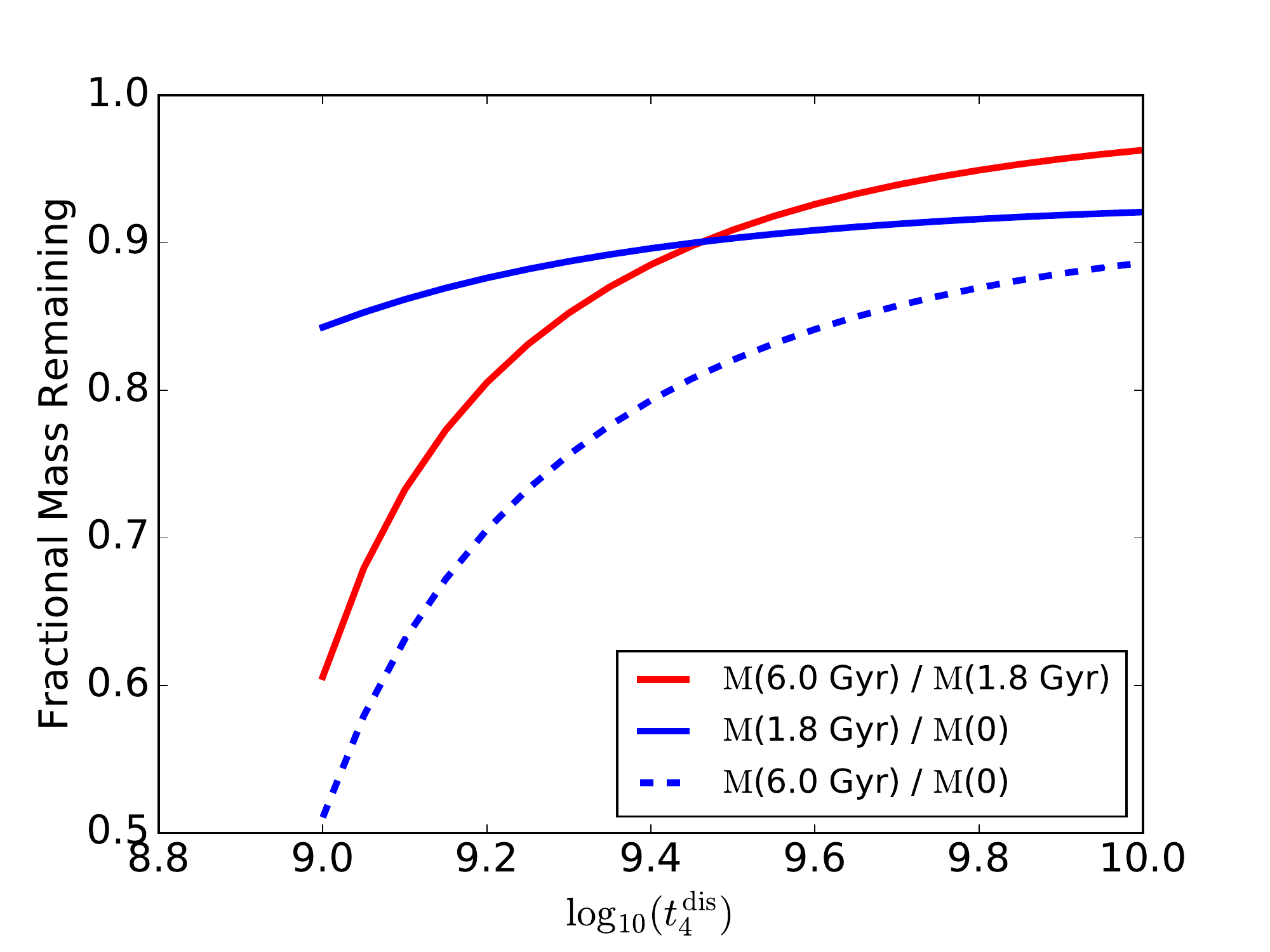}
\caption{Fractional mass remaining for NGC 1783 from its formation to
  ages of 1.8 Gyr and 6.0 Gyr (blue curves), and from 1.8 Gyr to 6.0
  Gyr (red).}
\label{fractional_mass_remaining}
\end{figure}

To quantitatively account for the amount of mass loss during the
evolution of a star cluster like NGC 1783, we calculated the
fractional mass loss expected for NGC 1783 from its present age of 1.8
Gyr to a more advanced age of 6 Gyr following \cite[][their equations
  1, 2, 3, and 6]{2005A&A...441..117L}. Unfortunately, the parameter
$t^{\rm dis}_4$, i.e., the dissolution timescale of a cluster with a
mass of $10^4\ \msun$, is poorly constrained for the LMC, $t^{\rm
  dis}_4 > 1$ Gyr \citep{2008MNRAS.383.1103P}. Therefore, we
investigated a wide range of choices for $t^{\rm dis}_4$, from 1 to 10
Gyr. We followed the evolution of the cluster's initial mass to ages
of 1.8 Gyr and 6 Gyr; we then calculated the ratio of both values: see
Fig. \ref{fractional_mass_remaining}. The smallest remaining mass
fraction from 1.8 Gyr to 6.0 Gyr is 60\%, corresponding to a lowest
remaining mass of $1.08 \times 10^5\ \msun$ by the time the cluster
will have evolved to an age of 6 Gyr. Note that
\citep{2017MNRAS.467.1857B} suggest that NGC 1783 may be capable of
retaining a significant fraction of its gas component (e.g., gas
generated by AGB ejecta), thus rendering our estimate a lower limit.

If we assume mass to be an important driver for the formation of MSPs,
then given our conclusion that the remaining mass of NGC 1783 at 6 Gyr
is likely at least $1.08 \times 10^5\ \msun$, we should expect to find
no evidence of chemical abundance variations in other, similarly
massive clusters with ages of around 6 Gyr. However, MSPs seem to be
ubiquitous for GCs older than 6 Gyr \citep[][their
  Fig. 23]{2017MNRAS.464.3636M}). For instance, NGC 339---an SMC
cluster with an age of 6.3 Gyr and a mass of $0.83 \times
10^5\ \msun$---has been reported to host MSPs \citep[][their
  Fig. 4]{2017MNRAS.465.4159N}. Therefore, a comparison of NGC 1783
with NGC 339 (and clusters with similar properties) suggests that the
idea of mass as a primary driver is not borne out by the
data. Although our result is consistent with age being a primary
factor \citep{2018MNRAS.473.2688M}, NGC 1783 only represents a single
data point so that we caution against overinterpretation of these
results.

Finally, the physical reality at the old age of genuine GCs may well
be rather different from that at the younger ages of the cluster we
have just investigated. \citep{2017MNRAS.464.3636M} suggested that
mass may be an important driver of MSPs in a sample of old Milky Way
GCs. They found clear correlations of both the fractional number and
the RGB pseudo-color width of the second-generation stars with their
host clusters' masses. Note, however, that chemical abundance
variations are only well established for old GCs; at the younger ages
discussed in this paper, we only have circumstantial evidence of
possible MSPs based on the broadening of distinct CMD features, not of
clearly distinct sequences.

\section{Summary and Conclusions}

Using {\sl HST} photometry, we have compared the $C_{\mathrm{index}}$
dispersion of our observed RGB sample with an artificial SSP, and
found their widths to be fully consistent with one another. We adopted
a color index that is sensitive to N and He abundance variations, so
the observed consistency indicates the absence of N and He abundance
variations, two elements whose abundances tend to vary most in star
clusters in general.

Our results thus indicate that an SSP model is sufficient to explain
the RGB broadening in NGC 1783. We explored the possible nature of two
outliers in the RGB distribution and concluded that they are most
likely field stars.

An intermediate-age cluster with an age of 1.8 Gyr and a mass of $1.8
\times 10^5\ \msun$, NGC 1783 is comparably massive as other star
clusters that show clear evidence of MSPs. After incorporating
mass-loss recipes from an age of 1.8 Gyr to a more advanced age of 6
Gyr, NGC 1783 is expected to remain as massive as some other clusters
that host clear MSPs.

If we assume that mass is an important driver of MSP formation, then
NGC 1783 should have shown clear evidence of chemical abundance
variations. However, we found no such evidence in NGC 1783. This
result is in agreement with the properties of other massive clusters
of similar age.

\section*{Acknowledgements}

This paper is based on observations made with the NASA/ESA {\sl HST},
and obtained from the Hubble Legacy Archive, which is a collaboration
of the Space Telescope Science Institute (STScI/NASA), the Space
Telescope European Coordinating Facility (ST-ECF/ESA), and the
Canadian Astronomy Data Centre (CADC/NRC/CSA). This work was supported
by the National Key Research and Development Program of China through
grant 2017YFA0402702. We also acknowledge research support from the
National Natural Science Foundation of China (grants U1631102,
11373010, and 11633005). H. Z. acknowledges support from the Chinese
National Innovation Training Program. C. L. acknowledges funding
support from the Macquarie Research Fellowship Scheme.

\bibliographystyle{aasjournal}
\bibliography{bibliography.bib}

\begin{thebibliography}{}
\expandafter\ifx\csname natexlab\endcsname\relax\def\natexlab#1{#1}\fi
\providecommand{\url}[1]{\href{#1}{#1}}

\bibitem[{{Anderson} {et~al.}(2009){Anderson}, {Piotto}, {King}, {Bedin}, \&
  {Guhathakurta}}]{2009ApJ...697L..58A}
{Anderson}, J., {Piotto}, G., {King}, I.~R., {Bedin}, L.~R., \& {Guhathakurta},
  P. 2009, \apjl, 697, L58

\bibitem[{{Balbinot} {et~al.}(2010){Balbinot}, {Santiago}, {Kerber}, {Barbuy},
  \& {Dias}}]{2010MNRAS.404.1625B}
{Balbinot}, E., {Santiago}, B.~X., {Kerber}, L.~O., {Barbuy}, B., \& {Dias},
  B.~M.~S. 2010, \mnras, 404, 1625

\bibitem[{{Bastian} {et~al.}(2013){Bastian}, {Cabrera-Ziri}, {Davies}, \&
  {Larsen}}]{2013MNRAS.436.2852B}
{Bastian}, N., {Cabrera-Ziri}, I., {Davies}, B., \& {Larsen}, S.~S. 2013,
  \mnras, 436, 2852

\bibitem[{{Bastian} {et~al.}(2015){Bastian}, {Cabrera-Ziri}, \&
  {Salaris}}]{2015MNRAS.449.3333B}
{Bastian}, N., {Cabrera-Ziri}, I., \& {Salaris}, M. 2015, \mnras, 449, 3333

\bibitem[{{Bastian} \& {Goodwin}(2006)}]{2006MNRAS.369L...9B}
{Bastian}, N., \& {Goodwin}, S.~P. 2006, \mnras, 369, L9

\bibitem[{{Bastian} \& {Strader}(2014)}]{2014MNRAS.443.3594B}
{Bastian}, N., \& {Strader}, J. 2014, \mnras, 443, 3594

\bibitem[{{Bate} {et~al.}(2003){Bate}, {Bonnell}, \&
  {Bromm}}]{2003MNRAS.339..577B}
{Bate}, M.~R., {Bonnell}, I.~A., \& {Bromm}, V. 2003, \mnras, 339, 577

\bibitem[{{Baumgardt} \& {Kroupa}(2007)}]{2007MNRAS.380.1589B}
{Baumgardt}, H., \& {Kroupa}, P. 2007, \mnras, 380, 1589

\bibitem[{{Baumgardt} {et~al.}(2013){Baumgardt}, {Parmentier}, {Anders}, \&
  {Grebel}}]{2013MNRAS.430..676B}
{Baumgardt}, H., {Parmentier}, G., {Anders}, P., \& {Grebel}, E.~K. 2013,
  \mnras, 430, 676

\bibitem[{{Bedin} {et~al.}(2000){Bedin}, {Piotto}, {Zoccali}, {Stetson},
  {Saviane}, {Cassisi}, \& {Bono}}]{2000A&A...363..159B}
{Bedin}, L.~R., {Piotto}, G., {Zoccali}, M., {et~al.} 2000, \aap, 363, 159

\bibitem[{{Bekki}(2017)}]{2017MNRAS.467.1857B}
{Bekki}, K. 2017, \mnras, 467, 1857

\bibitem[{{Bekki} \& {Mackey}(2009)}]{2009MNRAS.394..124B}
{Bekki}, K., \& {Mackey}, A.~D. 2009, \mnras, 394, 124

\bibitem[{{Bressan} {et~al.}(2012){Bressan}, {Marigo}, {Girardi}, {Salasnich},
  {Dal Cero}, {Rubele}, \& {Nanni}}]{2012MNRAS.427..127B}
{Bressan}, A., {Marigo}, P., {Girardi}, L., {et~al.} 2012, \mnras, 427, 127

\bibitem[{{Cabrera-Ziri} {et~al.}(2014){Cabrera-Ziri}, {Bastian}, {Davies},
  {Magris}, {Bruzual}, \& {Schweizer}}]{2014MNRAS.441.2754C}
{Cabrera-Ziri}, I., {Bastian}, N., {Davies}, B., {et~al.} 2014, \mnras, 441,
  2754

\bibitem[{{Cabrera-Ziri} {et~al.}(2016){Cabrera-Ziri}, {Lardo}, {Davies},
  {Bastian}, {Beccari}, {Larsen}, \& {Hernandez}}]{2016MNRAS.460.1869C}
{Cabrera-Ziri}, I., {Lardo}, C., {Davies}, B., {et~al.} 2016, \mnras, 460, 1869

\bibitem[{{Cabrera-Ziri} {et~al.}(2015){Cabrera-Ziri}, {Bastian}, {Longmore},
  {Brogan}, {Hollyhead}, {Larsen}, {Whitmore}, {Johnson}, {Chandar}, {Henshaw},
  {Davies}, \& {Hibbard}}]{2015MNRAS.448.2224C}
{Cabrera-Ziri}, I., {Bastian}, N., {Longmore}, S.~N., {et~al.} 2015, \mnras,
  448, 2224

\bibitem[{{Carretta} {et~al.}(2004){Carretta}, {Bragaglia}, \&
  {Cacciari}}]{2004ApJ...610L..25C}
{Carretta}, E., {Bragaglia}, A., \& {Cacciari}, C. 2004, \apjl, 610, L25

\bibitem[{{Carretta} {et~al.}(2009){Carretta}, {Bragaglia}, {Gratton},
  {Lucatello}, {Catanzaro}, {Leone}, {Bellazzini}, {Claudi}, {D'Orazi},
  {Momany}, {Ortolani}, {Pancino}, {Piotto}, {Recio-Blanco}, \&
  {Sabbi}}]{2009A&A...505..117C}
{Carretta}, E., {Bragaglia}, A., {Gratton}, R.~G., {et~al.} 2009, \aap, 505,
  117

\bibitem[{{Charbonnel} {et~al.}(2013){Charbonnel}, {Krause}, {Decressin},
  {Prantzos}, \& {Meynet}}]{2013MmSAI..84..158C}
{Charbonnel}, C., {Krause}, M., {Decressin}, T., {Prantzos}, N., \& {Meynet},
  G. 2013, \memsai, 84, 158

\bibitem[{{Davidge}(2012)}]{2012ApJ...761..155D}
{Davidge}, T.~J. 2012, \apj, 761, 155

\bibitem[{{Denissenkov} \& {Hartwick}(2014)}]{2014MNRAS.437L..21D}
{Denissenkov}, P.~A., \& {Hartwick}, F.~D.~A. 2014, \mnras, 437, L21

\bibitem[{{D'Ercole} {et~al.}(2008){D'Ercole}, {Vesperini}, {D'Antona},
  {McMillan}, \& {Recchi}}]{2008MNRAS.391..825D}
{D'Ercole}, A., {Vesperini}, E., {D'Antona}, F., {McMillan}, S.~L.~W., \&
  {Recchi}, S. 2008, \mnras, 391, 825

\bibitem[{{Dolphin}(2000)}]{2000PASP..112.1383D}
{Dolphin}, A.~E. 2000, \pasp, 112, 1383

\bibitem[{{Farias} {et~al.}(2015){Farias}, {Smith}, {Fellhauer}, {Goodwin},
  {Candlish}, {Bla{\~n}a}, \& {Dominguez}}]{2015MNRAS.450.2451F}
{Farias}, J.~P., {Smith}, R., {Fellhauer}, M., {et~al.} 2015, \mnras, 450, 2451

\bibitem[{{Geisler} {et~al.}(1997){Geisler}, {Bica}, {Dottori}, {Claria},
  {Piatti}, \& {Santos}}]{1997AJ....114.1920G}
{Geisler}, D., {Bica}, E., {Dottori}, H., {et~al.} 1997, \aj, 114, 1920

\bibitem[{{Goudfrooij} {et~al.}(2011){Goudfrooij}, {Puzia},
  {Kozhurina-Platais}, \& {Chandar}}]{2011ApJ...737....3G}
{Goudfrooij}, P., {Puzia}, T.~H., {Kozhurina-Platais}, V., \& {Chandar}, R.
  2011, \apj, 737, 3

\bibitem[{{Gratton} {et~al.}(2004){Gratton}, {Sneden}, \&
  {Carretta}}]{2004ARA&A..42..385G}
{Gratton}, R., {Sneden}, C., \& {Carretta}, E. 2004, \araa, 42, 385

\bibitem[{{Harris}(1996)}]{1996AJ....112.1487H}
{Harris}, W.~E. 1996, \aj, 112, 1487

\bibitem[{{Hong} {et~al.}(2017){Hong}, {de Grijs}, {Askar}, {Berczik}, {Li},
  {Wang}, {Deng}, {Kouwenhoven}, {Giersz}, \& {Spurzem}}]{2017MNRAS.472...67H}
{Hong}, J., {de Grijs}, R., {Askar}, A., {et~al.} 2017, \mnras, 472, 67

\bibitem[{{Jiang} {et~al.}(2014){Jiang}, {Han}, \& {Li}}]{2014ApJ...789...88J}
{Jiang}, D., {Han}, Z., \& {Li}, L. 2014, \apj, 789, 88

\bibitem[{{Lamers} {et~al.}(2005){Lamers}, {Gieles}, {Bastian}, {Baumgardt},
  {Kharchenko}, \& {Portegies Zwart}}]{2005A&A...441..117L}
{Lamers}, H.~J.~G.~L.~M., {Gieles}, M., {Bastian}, N., {et~al.} 2005, \aap,
  441, 117

\bibitem[{{Li} {et~al.}(2016{\natexlab{a}}){Li}, {de Grijs}, {Deng}, {Geller},
  {Xin}, {Hu}, \& {Faucher-Gigu{\`e}re}}]{2016Natur.529..502L}
{Li}, C., {de Grijs}, R., {Deng}, L., {et~al.} 2016{\natexlab{a}}, \nat, 529,
  502

\bibitem[{{Li} {et~al.}(2016{\natexlab{b}}){Li}, {de Grijs}, \&
  {Deng}}]{2016RAA....16..179L}
{Li}, C.-Y., {de Grijs}, R., \& {Deng}, L.-C. 2016{\natexlab{b}}, Research in
  Astronomy and Astrophysics, 16, 179

\bibitem[{{Lima} {et~al.}(2014){Lima}, {Bica}, {Bonatto}, \&
  {Saito}}]{2014A&A...568A..16L}
{Lima}, E.~F., {Bica}, E., {Bonatto}, C., \& {Saito}, R.~K. 2014, \aap, 568,
  A16

\bibitem[{{Mackey} \& {Broby Nielsen}(2007)}]{2007MNRAS.379..151M}
{Mackey}, A.~D., \& {Broby Nielsen}, P. 2007, \mnras, 379, 151

\bibitem[{{Marino} {et~al.}(2008){Marino}, {Villanova}, {Piotto}, {Milone},
  {Momany}, {Bedin}, \& {Medling}}]{2008A&A...490..625M}
{Marino}, A.~F., {Villanova}, S., {Piotto}, G., {et~al.} 2008, \aap, 490, 625

\bibitem[{{Martocchia} {et~al.}(2017){Martocchia}, {Bastian}, {Usher},
  {Kozhurina-Platais}, {Niederhofer}, {Cabrera-Ziri}, {Dalessandro},
  {Hollyhead}, {Kacharov}, {Lardo}, {Larsen}, {Mucciarelli}, {Platais},
  {Salaris}, {Cordero}, {Geisler}, {Hilker}, {Li}, \&
  {Mackey}}]{2017MNRAS.468.3150M}
{Martocchia}, S., {Bastian}, N., {Usher}, C., {et~al.} 2017, \mnras, 468, 3150

\bibitem[{{Martocchia} {et~al.}(2018){Martocchia}, {Cabrera-Ziri}, {Lardo},
  {Dalessandro}, {Bastian}, {Kozhurina-Platais}, {Usher}, {Niederhofer},
  {Cordero}, {Geisler}, {Hollyhead}, {Kacharov}, {Larsen}, {Li}, {Mackey},
  {Hilker}, {Mucciarelli}, {Platais}, \& {Salaris}}]{2018MNRAS.473.2688M}
{Martocchia}, S., {Cabrera-Ziri}, I., {Lardo}, C., {et~al.} 2018, \mnras, 473,
  2688

\bibitem[{{McLaughlin} \& {van der Marel}(2005)}]{2005ApJS..161..304M}
{McLaughlin}, D.~E., \& {van der Marel}, R.~P. 2005, \apjs, 161, 304

\bibitem[{{Milone} {et~al.}(2009){Milone}, {Bedin}, {Piotto}, \&
  {Anderson}}]{2009A&A...497..755M}
{Milone}, A.~P., {Bedin}, L.~R., {Piotto}, G., \& {Anderson}, J. 2009, \aap,
  497, 755

\bibitem[{{Milone} {et~al.}(2012{\natexlab{a}}){Milone}, {Piotto}, {Bedin},
  {King}, {Anderson}, {Marino}, {Bellini}, {Gratton}, {Renzini}, {Stetson},
  {Cassisi}, {Aparicio}, {Bragaglia}, {Carretta}, {D'Antona}, {Di Criscienzo},
  {Lucatello}, {Monelli}, \& {Pietrinferni}}]{2012ApJ...744...58M}
{Milone}, A.~P., {Piotto}, G., {Bedin}, L.~R., {et~al.} 2012{\natexlab{a}},
  \apj, 744, 58

\bibitem[{{Milone} {et~al.}(2012{\natexlab{b}}){Milone}, {Piotto}, {Bedin},
  {Aparicio}, {Anderson}, {Sarajedini}, {Marino}, {Moretti}, {Davies},
  {Chaboyer}, {Dotter}, {Hempel}, {Mar{\'{\i}}n-Franch}, {Majewski}, {Paust},
  {Reid}, {Rosenberg}, \& {Siegel}}]{2012A&A...540A..16M}
---. 2012{\natexlab{b}}, \aap, 540, A16

\bibitem[{{Milone} {et~al.}(2015){Milone}, {Marino}, {Piotto}, {Bedin},
  {Anderson}, {Renzini}, {King}, {Bellini}, {Brown}, {Cassisi}, {D'Antona},
  {Jerjen}, {Nardiello}, {Salaris}, {Marel}, {Vesperini}, {Yong}, {Aparicio},
  {Sarajedini}, \& {Zoccali}}]{2015MNRAS.447..927M}
{Milone}, A.~P., {Marino}, A.~F., {Piotto}, G., {et~al.} 2015, \mnras, 447, 927

\bibitem[{{Milone} {et~al.}(2017){Milone}, {Piotto}, {Renzini}, {Marino},
  {Bedin}, {Vesperini}, {D'Antona}, {Nardiello}, {Anderson}, {King}, {Yong},
  {Bellini}, {Aparicio}, {Barbuy}, {Brown}, {Cassisi}, {Ortolani}, {Salaris},
  {Sarajedini}, \& {van der Marel}}]{2017MNRAS.464.3636M}
{Milone}, A.~P., {Piotto}, G., {Renzini}, A., {et~al.} 2017, \mnras, 464, 3636

\bibitem[{{Mucciarelli} {et~al.}(2008){Mucciarelli}, {Carretta}, {Origlia}, \&
  {Ferraro}}]{2008AJ....136..375M}
{Mucciarelli}, A., {Carretta}, E., {Origlia}, L., \& {Ferraro}, F.~R. 2008,
  \aj, 136, 375

\bibitem[{{Niederhofer} {et~al.}(2017{\natexlab{a}}){Niederhofer}, {Bastian},
  {Kozhurina-Platais}, {Larsen}, {Hollyhead}, {Lardo}, {Cabrera-Ziri},
  {Kacharov}, {Platais}, {Salaris}, {Cordero}, {Dalessandro}, {Geisler},
  {Hilker}, {Li}, {Mackey}, \& {Mucciarelli}}]{2017MNRAS.465.4159N}
{Niederhofer}, F., {Bastian}, N., {Kozhurina-Platais}, V., {et~al.}
  2017{\natexlab{a}}, \mnras, 465, 4159

\bibitem[{{Niederhofer} {et~al.}(2017{\natexlab{b}}){Niederhofer}, {Bastian},
  {Kozhurina-Platais}, {Larsen}, {Salaris}, {Dalessandro}, {Mucciarelli},
  {Cabrera-Ziri}, {Cordero}, {Geisler}, {Hilker}, {Hollyhead}, {Kacharov},
  {Lardo}, {Li}, {Mackey}, \& {Platais}}]{2017MNRAS.464...94N}
---. 2017{\natexlab{b}}, \mnras, 464, 94

\bibitem[{{Parmentier} \& {de Grijs}(2008)}]{2008MNRAS.383.1103P}
{Parmentier}, G., \& {de Grijs}, R. 2008, \mnras, 383, 1103

\bibitem[{{Pflamm-Altenburg} \& {Kroupa}(2009)}]{2009MNRAS.397..488P}
{Pflamm-Altenburg}, J., \& {Kroupa}, P. 2009, \mnras, 397, 488

\bibitem[{{Piotto} {et~al.}(2013){Piotto}, {Milone}, {Marino}, {Bedin},
  {Anderson}, {Jerjen}, {Bellini}, \& {Cassisi}}]{2013ApJ...775...15P}
{Piotto}, G., {Milone}, A.~P., {Marino}, A.~F., {et~al.} 2013, \apj, 775, 15

\bibitem[{{Piotto} {et~al.}(2007){Piotto}, {Bedin}, {Anderson}, {King},
  {Cassisi}, {Milone}, {Villanova}, {Pietrinferni}, \&
  {Renzini}}]{2007ApJ...661L..53P}
{Piotto}, G., {Bedin}, L.~R., {Anderson}, J., {et~al.} 2007, \apjl, 661, L53

\bibitem[{{Piotto} {et~al.}(2015){Piotto}, {Milone}, {Bedin}, {Anderson},
  {King}, {Marino}, {Nardiello}, {Aparicio}, {Barbuy}, {Bellini}, {Brown},
  {Cassisi}, {Cool}, {Cunial}, {Dalessandro}, {D'Antona}, {Ferraro}, {Hidalgo},
  {Lanzoni}, {Monelli}, {Ortolani}, {Renzini}, {Salaris}, {Sarajedini}, {van
  der Marel}, {Vesperini}, \& {Zoccali}}]{2015AJ....149...91P}
{Piotto}, G., {Milone}, A.~P., {Bedin}, L.~R., {et~al.} 2015, \aj, 149, 91

\bibitem[{{Rubele} {et~al.}(2013){Rubele}, {Girardi}, {Kozhurina-Platais},
  {Kerber}, {Goudfrooij}, {Bressan}, \& {Marigo}}]{2013MNRAS.430.2774R}
{Rubele}, S., {Girardi}, L., {Kozhurina-Platais}, V., {et~al.} 2013, \mnras,
  430, 2774

\bibitem[{{Shetrone}(1996)}]{1996AJ....112.1517S}
{Shetrone}, M.~D. 1996, \aj, 112, 1517

\bibitem[{{Stetson}(1987)}]{1987PASP...99..191S}
{Stetson}, P.~B. 1987, \pasp, 99, 191

\bibitem[{{Vesperini} {et~al.}(2013){Vesperini}, {McMillan}, {D'Antona}, \&
  {D'Ercole}}]{2013MNRAS.429.1913V}
{Vesperini}, E., {McMillan}, S.~L.~W., {D'Antona}, F., \& {D'Ercole}, A. 2013,
  \mnras, 429, 1913

\bibitem[{{Worthey}(1994)}]{1994ApJS...95..107W}
{Worthey}, G. 1994, \apjs, 95, 107

\bibitem[{{Wu} {et~al.}(2016){Wu}, {Li}, {de Grijs}, \&
  {Deng}}]{2016ApJ...826L..14W}
{Wu}, X., {Li}, C., {de Grijs}, R., \& {Deng}, L. 2016, \apjl, 826, L14

\end{thebibliography}

\end{document}